\documentclass[a4paper,10pt]{article}
\usepackage[utf8]{inputenc}

\title{Violation of the Holographic Principle in the Loop Quantum Gravity}
\author{Ozan Sarg{\i}n$^1$ and   Mir Faizal$^2$ \\ \\
$^1$Department of Physics,\\
Izmir Institute of Technology,\\ TR35430, Izmir, Turkey\\ \\
$^2$Department of Physics and Astronomy,\\ University of Waterloo, \\Waterloo, ON, N2L 3G1, Canada.
}
\date{}
\begin{document}

\maketitle

\begin{abstract}
In this paper, we analyze the holographic principle using loop quantum gravity (LQG).  
This will be done by using  polymeric quantization for analysing the Yurtsever's holographic  bound on 
the entropy, which is  obtained from local quantum field theories.  
As the polymeric quantization is the characteristic feature of loop quantum gravity, we will argue that this 
calculation will indicate the effect of  loop quantum gravity on the  holographic principle. 
Thus, we will be able to explicitly demonstrate    the violation of the holographic principle  in    the loop quantum gravity. 
\end{abstract}

\section{Introduction}
Loop quantum gravity (LQG) is a background independent, nonperturbative approach to quantize Einstein's 
General Theory of Relativity \cite{lqg1, lqg2}. It is one of the main approaches to the 
quantum theory of gravity, and so  LQG necessitates a quantized structure for 
space. The microscopic degrees of freedom in LQG are discrete, and it is important to relate these discrete degrees 
of freedom to the macroscopic geometry which is described by a continuous differential manifold. 
Thus, the LQG is constrained to  produce a  continuous differential manifold in the infrared limit. 
This has been done by using polymeric quantization, which is characterized by a polymeric 
length scale \cite{pq, pq1}. At
scales much larger than the polymeric length scale, the geometry of spacetime can be approximated by a
differential manifold. However, at scales comparable to polymeric length scale, the discrete behavior 
becomes more manifest. This has also led to the development of polymer quantum mechanics. 
So, polymeric quantum mechanics can be used to understand the behavior of LQG, as it 
is the main  technique  used in loop quantum gravity. It may be noted that quantum field theory  
consistent with    polymeric quantization has been  studied \cite{qf}.
This has been done by analysing  the   propagator of  scalar field theory using a method which took 
the polymeric effects into consideration. 
Thus, an expression for the propagator, consistent with polymeric quantization,  was obtained both 
  in the infrared and the ultraviolet regimes.  The semi-classical dynamics of a 
  free massive scalar field in a homogeneous and isotropic cosmological spacetime has been used for 
  analyzing polymeric inflation \cite{in, in12}.

In this paper, we will analyse the effects of the polymeric quantization on the holographic principle. 
The 
holographic principle states that   degrees of freedom in a region in space is equal to 
 degrees of freedom on the boundary  surrounding that region of space \cite{holo, holo1}. 
 The t' Hooft's holographic principle  has been motivated from the  
Bekenstein entropy-area relation,  as according to this relation the entropy of the black hole scales with 
its area, $S = A/4$ \cite{z1,jd, 4a, z2}. 
The Bekenstein entropy-area relation  is expected to  get modified  
at very small scales  where   quantum gravitational effects become significant
\cite{l1, card, other, other0, other1,  solo1, solo4, solo5,jy, bss}. 
So, it is expected that the holographic principle will get violated near Planck scale 
\cite{viol, viol1}.

A  different kind of holographic bound  on entropy  called the Yurtsever's holographic bound  occurs in local quantum field theories 
  \cite{a, b,c,d}.
  Unlike the t' Hooft's holographic bound   
  where the entropy scales as $A$ \cite{holo1},
  in the Yurtsever's holographic bound the entropy scales as $A^{3/4}$ \cite{a, b,c,d}.
  The Yurtsever’s holographic bound is obtained by    imposing an upper bound on the total energy of the
corresponding Fock states which ensures that the system is in a stable configuration against
gravitational collapse,  and imposing a cutoff on the maximum energy of the field modes of the
order of   Planck energy.  
It is this bound that will be violated  
by polymeric quantization because the polymeric quantization effectively modifies the measure of the phase space. 
It may be noted that it has been demonstrated that this bound also gets   violated due to the existence of  minimum measurable length scale 
because such a minimum measurable length scale   modifies the usual uncertainty principle to a 
generalized uncertainty principle, and this generalized uncertainty principle in turn modifies the measure of the phase space 
\cite{ali}.  As the generalized uncertainty principle is related to the polymeric quantization \cite{ba},
it was   expected that 
a similar result will also hold for polymeric quantization. However, this was never explicitly demonstrated, and this is what we  to do in this 
paper. Furthermore, as the loop quantum gravity depends on polymeric quantization, 
our calculations indicate a violation of the        holographic  principle in loop quantum gravity.

It may be noted that   the area of the black hole increases by a Planck area when   one bit of information crosses 
the horizon \cite{jd}. 
So,  even using the usual black hole thermodynamics it can be demonstrated that 
the t' Hooft's holographic entropy bound  is an integer multiple  of the Planck area. This restricts the number 
of microstates for black hole in any theory of quantum gravity. 
In fact, it has been explicitly demonstrated that the  entropy of black hole is an integer multiple of the Planck area using the 
 spacetime foam picture, which is based on  the usual Wheeler-DeWitt approach \cite{r1, r2}.
So, discreteness of space cannot increase the t' Hooft's holographic bound on the entropy, as this discreteness   
cannot modify the number of microstates for a black hole. It can only  explain the nature of such 
 microstates.    
However, the discretization of space in polymeric quantization  
effectively changes the measure used in the  
  phase space. This can   increase    the Yurtsever's holographic  bound on the entropy in local quantum field theories, 
  and this is what we will explicitly demonstrate in this paper.

 \section{Polymer Quantization}
In this section, we will review the polymer quantization \cite{pq, pq1, pq3}. 
Polymer quantization is a singular representation of the Weyl-Heisenberg algebra. It is  
unitarity inequivalent to the familiar Schr\"{o}dinger representation. In order to implement this 
  quantization scheme, a representation for the algebra of this theory has to be 
  constructed. So, we have to obtain a suitable 
  inner product to obtain the  kinematic Hilbert space. 
  After defining a suitable inner product, the dynamics of this theory can be analysed. This can be done by 
  first noting that the   polymer Hilbert space is a non-separable space which is defined as 
$H_{poly}=L^{2}(\mathcal{R}_{d},d_{\mu_{d}})$. Thus, 
polymer Hilbert space is the Cauchy completion of the space 
of complex-valued  functions which are cylindrical with respect to a graph on the real line. 
Now if $|x_{i}\rangle$ are the normalizable eigenvectors of the position 
operator which span $H_{poly}$, then we can write an 
 inner product  as 
\begin{eqnarray}\label{inner}
  \langle x_{i}|x_{j}\rangle=\delta_{i,j}.  
\end{eqnarray}
This inner product  can be used to obtain the 
  Cauchy completion of the cylindrical function space.
The position and momentum operators can not be defined simultaneously. 
This is because the states in $H_{poly}$ have support on graphs which have countably many points in them. 
Thus, we cannot define   the momentum operator through differentiation.
However, the Hamiltonian of a system contains the kinetic terms, and so we need to have a representation 
for the momentum operator. Thus, the momentum operator is defined by 
 regularizing the Hamiltonian. This is done  by introducing a lattice structure 
on the position space,  and then using it to define a translation operator. 
This translation operator  replaces the momentum operator in polymeric quantization.  
Now we can write the    operators  which  are commonly used    in polymer 
quantization, 
\begin{eqnarray}
\label{ops}
  \hat{x}|x_{j}\rangle &=& x_{j}|x_{j}\rangle \\
  \hat{V}(\mu)|x_{j}\rangle &=& |x_{j}-\mu\rangle
\end{eqnarray}
where $\mu$ is the regulation measure. This regulation measure cannot be 
  removed from the results of 
the theory. It may be noted that this regulation measure
has   ambiguity related to its exact value. Now the momentum operator can be written as 
\begin{eqnarray}\label{mom}
  \hat{p}=\frac{\hat{V}(\mu)-\hat{V^{\dag}}(\mu)}{2i\mu}\;.
\end{eqnarray}
In the limit $\mu \to 0$, this operator reduces to  $\hat{p}=-i\frac{\partial}{\partial_{x}}$, 
which is the usual   momentum operator. Even though this  polymerization measure $\mu$ depends on the position, 
in the semi-classical approximation, a constant $\mu$ is used. Furthermore, the   translation 
operator is replaced by    $\hat{V}(\mu)\rightarrow e^{i\mu \hat{p}}$. Thus, the momentum 
operator can be written as  $\hat{p}_{\mu}=\frac{\sin(\mu p)}{\mu}$. 

Now let us consider the effect of polymer quantization on a one-dimensional system,  with the Hamiltonian  
\begin{eqnarray}\label{1-d hamiltonian}
H=\frac{p^{2}_{\mu}}{2m}+V(x),
\end{eqnarray}
where we have used the polymer modified momentum 
$p_{\mu}=\frac{\sin(\mu p)}{\mu}$ instead of the canonical one.  
The modified equations of motion  now take the form \cite{Liou}, 
\begin{eqnarray}\label{eqn_of motion1}
  \dot{x}=\{x,H\}=\{x,p_{\mu}\}\frac{\partial H}{\partial p_{\mu}}=\cos(\mu p)\frac{p_{\mu}}{m}=\sqrt{1-\mu^{2}p_{\mu}^{2}}\;\frac{p_{\mu}}{m}
  \\ \label{eqn_of motion2}
  \dot{p}_{\mu}=\{p_{\mu},H\}=-\{x,p_{\mu}\}\frac{\partial H}{\partial x}=\cos(\mu p)\Big(-\frac{\partial V}{\partial x}\Big)=\sqrt{1-\mu^{2}p_{\mu}^{2}}\Big(-\frac{\partial V}{\partial x}\Big).
\end{eqnarray}
Using (\ref{eqn_of motion1}) and (\ref{eqn_of motion2}), we can write   the acceleration as 
\begin{eqnarray}\label{acceleration5}
  \ddot{x}=\frac{1}{m}\Big(-\frac{\partial V}{\partial x}\Big)\Big[1-2\mu ^{2}p^{2}_{\mu}\Big].
\end{eqnarray}
In this section, we reviewed some basic results relating to polymer quantization. 
  In the next section, we will apply the polymer quantization 
to the Liouville theorem.  

\section{The Polymer Quantization and Liouville Theorem}
It is possible to analyse the effect of the 
polymeric quantization on  the   Liouville theorem.  
The number of states inside a volume of phase space   does not
change with time, the time variations of position and momentum are   $  x_{i}^{'}=  x_{i}+\delta x_{i}$ and $  p_{\mu_{\,i}}^{'}= p_{\mu_{\,i}}+\delta p_{\mu_{\,i}}$.  
Using (\ref{eqn_of motion1}) and (\ref{eqn_of motion2}) the infinitesimal changes in position and momentum become
\begin{eqnarray}\label{delta_x}
  \delta x_{i}&=&\{x_{i},p_{\mu_{\,j}}\}\frac{\partial H}{\partial p_{\mu_{\,j}}}\delta t
\nonumber \\ 
  \delta p_{\mu_{\,i}}&=&-\{x_{j},p_{\mu_{\,i}}\}\frac{\partial H}{\partial x_{j}}\delta t\:.
\end{eqnarray}
The infinitesimal phase space volume after this infinitesimal time evolution can now be written as 
\begin{eqnarray}\label{Liou1}
  d^{D}x^{'} d^{D}p_{\mu}^{'}=\Bigg|\frac{\partial(x_{1}^{'},\ldots,x_{D}^{'},p_{\mu_{\,1}}^{'},
  \ldots,p_{\mu_{\,D}}^{'})}{\partial(x_{1},\ldots,x_{D},p_{\mu_{\,1}},\ldots,p_{\mu_{\,D}})}\Bigg|d^{D}x\, d^{D}p_{\mu}
\end{eqnarray}
where 
\begin{eqnarray}\label{Jacob}
 \Bigg|\frac{\partial(x_{1}^{'},\ldots,x_{D}^{'},p_{\mu_{\,1}}^{'},
  \ldots,p_{\mu_{\,D}}^{'})}{\partial(x_{1},\ldots,x_{D},p_{\mu_{\,1}},\ldots,p_{\mu_{\,D}})}\Bigg|&=& 1+\Bigg(\frac{\partial\delta x_{i}}{\partial x_{i}}+\frac{\partial\delta p_{\mu_{\,i}}}{\partial p_{\mu_{\,i}}}\Bigg)+\cdots
\nonumber \\ 
  \Bigg(\frac{\partial\delta x_{i}}{\partial x_{i}}+\frac{\partial\delta p_{\mu_{\,i}}}{\partial p_{\mu_{\,i}}}\Bigg)
  \frac{1}{\delta t}&=&-\Bigg[\frac{\partial}{\partial p_{\mu_{\,i}}}\{x_{j},p_{\mu_{\,i}}\}\Bigg]\frac{\partial H}{\partial x_{j}}\:.
\end{eqnarray}
We know that 
\begin{eqnarray}\label{poisson1}
  \{x_{i},p_{\mu_{\,j}}\}&=&\delta_{ij}\sqrt{1-\mu^{2}p_{\mu_{\,j}}^{\:2}}
\nonumber \\ 
  -\Bigg[\frac{\partial}{\partial p_{\mu_{\,i}}}\{x_{j},p_{\mu_{\,i}}\}\Bigg]&=& \delta_{ij}\mu^{2}p_{\mu_{\,i}}\:.
\end{eqnarray}
Now we can write (\ref{Liou1}) as 
\begin{eqnarray}\label{Liou5}
  d^{D}x^{'} d^{D}p_{\mu}^{'}=\Bigg(1+\mu ^{2}p_{\mu_{\,i}}\frac{\partial H}{\partial x_{i}}\delta t\Bigg)d^{D}x\, d^{D}p_{\mu}\:.
\end{eqnarray}
We have to get rid of the term in the parenthesis  in order to cast (\ref{Liou5})  into an invariant form. To this end, let's consider the infinitesimal evolution of $(1-\mu^{2}{p_{\mu}^{'}}^{2})$, up to the 
first order in $\mu^{2}$ and $\delta t$. It is given by 
\begin{eqnarray}\label{invariant1}
  \Big(1-\mu^{2}{p_{\mu}^{'}}^{2}\Big)&=& 1-\mu^{2}\Bigg(p_{\mu_{\,i}}-\sqrt{1-\mu^{2}p_{\mu_{\,i}}^{\:2}}\frac{\partial H}{\partial x_{i}}\delta t\Bigg)^{2}
\end{eqnarray}
where we have used the fact that 
\begin{eqnarray}\label{p_evol1}
  p_{\mu_{\,i}}^{'}= p_{\mu_{\,i}}+\delta p_{\mu_{\,i}}=p_{\mu_{\,i}}-\sqrt{1-\mu^{2}p_{\mu_{\,i}}^{\:2}}\:\Big(\frac{\partial H}{\partial x_{i}}\Big)\delta t\:.
\end{eqnarray}
So, after a little bit of algebra, we can write 
\begin{eqnarray}\label{invariant5}
  \Big(1-\mu^{2}{p_{\mu}^{'}}^{2}\Big)=\Big(1-\mu^{2}{p_{\mu}}^{2}\Big)\Bigg[1+2\mu^{2}\,p_{\mu_{\,i}}\frac{\partial H}{\partial x_{i}}\delta t\Bigg]\;.
\end{eqnarray}
Thus, we obtain 
\begin{eqnarray}\label{invariant6}
  \Big(1-\mu^{2}{p_{\mu}^{'}}^{2}\Big)^{-\frac{1}{2}}=\Big(1-\mu^{2}{p_{\mu}}^{2}\Big)^{-\frac{1}{2}}\Bigg[1-\mu^{2}\,p_{\mu_{\,i}}\frac{\partial H}{\partial x_{i}}\delta t\Bigg]\:.
\end{eqnarray}
Now, if we multiply both sides of (\ref{Liou5}) by (\ref{invariant6}) 
we obtain, 
\begin{eqnarray}\label{Liou7}
  \frac{d^{D}x^{'} d^{D}p_{\mu}^{'}}{\sqrt{\Big(1-\mu^{2}{p_{\mu}^{'}}^{2}\Big)}}=\frac{d^{D}x\, d^{D}p_{\mu}}{\sqrt{\Big(1-\mu^{2}{p_{\mu}}^{2}\Big)}}\Bigg[1+\mu ^{2}p_{\mu_{\,i}}\frac{\partial H}{\partial x_{i}}\delta t\Bigg]\Bigg[1-\mu ^{2}p_{\mu_{\,i}}\frac{\partial H}{\partial x_{i}}\delta t\Bigg]
\end{eqnarray}
where,
\begin{eqnarray}\label{Liou8}
  \Bigg[1+\mu ^{2}p_{\mu_{\,i}}\frac{\partial H}{\partial x_{i}}\delta t\Bigg]\Bigg[1-\mu ^{2}p_{\mu_{\,i}}\frac{\partial H}{\partial x_{i}}\delta t\Bigg]
\approx 1\:.
\end{eqnarray}
Therefore the invariant volume element of the phase space takes the following form, 
\begin{eqnarray}\label{Liou9}
   \frac{d^{D}x^{'} d^{D}p_{\mu}^{'}}{\sqrt{\Big(1-\mu^{2}{p_{\mu}^{'}}^{2}\Big)}}=\frac{d^{D}x\, d^{D}p_{\mu}}{\sqrt{\Big(1-\mu^{2}{p_{\mu}}^{2}\Big)}}\:.
\end{eqnarray}
After integrating over the coordinates, the infinitesimal volume element of the phase space can be written as 
$  {V \;d^{D}p_{\mu}}/{\sqrt{ (1-\mu^{2}{p_{\mu}}^{2})}}$. 
The number of states per momentum space volume is assumed to be $  {V}
 {d^{D}p_{\mu}}/{(2\Pi\hbar)^{D}}{\sqrt{(1-\mu^{2}{p_{\mu}}^{2})}}$. The modification of the number of states per momentum, will 
 modify the entropy of a local quantum field theory. This will in turn violate the holographic bound. 
 
\section{The Yurtsever's Holographic Bound }
The holographic principle states that the degrees of freedom in a region of space is equal to 
the degrees of freedom on the boundary of that region of space. The holographic principle can be 
used to analyse a 
  closed spacelike surface containing quantum  bosonic fields
 \cite{a, b, c,d, ali}. 
  Now  at a  temperature $T$,
the energy of the most probable state for this field theory can be written as $E= a_1 Z T^4 V$
where $Z$ is the number of different fundamental particles   with mass less than $T$,  and
$a_1$ is a suitably chosen  constant. In natural units it can be taken to be of  order one.  
We can also write the entropy of this system as $S=a_2 Z V T^3$, where $a_2$ is another constant which can  again be chosen to be of order one. 
We obtain the holographic   limit  
$ 2 E <\frac{V}{\frac{4}{3} \pi}, $ and we also obtain $T<a_3 Z^{-\frac{1}{4}} V^{-\frac{1}{6}}$. 
So, the entropy bound is given by 
\begin{eqnarray}
 S<a_4  Z^{\frac{1}{4}} V^{\frac{1}{2}}= a_4  Z^{\frac{1}{4}} A^{\frac{3}{4}},
\end{eqnarray}
where $A $ is the area of the boundary. 
At low temperatures, Yurtsever's holographic bound   is small compared to t' Hooft's holographic bound, $S = A/4$ \cite{holo, holo1}. 
It may be noted that here  the bulk  degrees of freedom  for any closed surface are equal to the boundary degrees of freedom.
So, the physics inside a closed surface can be 
  be represented by surface degrees
of freedom.   

It is possible to derive the Yurtsever's holographic bound on the entropy  using local quantum field theories 
 in flat spacetime,  where the     total energy of Fock states is in a stable configuration \cite{a,b,c,d}.  
 This bound is obtained  by 
imposing a cutoff on the maximum energy that 
modes of the quantum field can attain. This cutoff is usually of order of Planck energy. So, 
the total number of the quantized modes for a massless bosonic field confined to  cubic
box of size $L$, can be written as 
\begin{eqnarray}
 N=\sum_{\vec{k}}1 \rightarrow \frac{L^3}{(2\pi)^3}\int d^3\vec{p}
  =\frac{L^3}{2\pi^2}\int_{0}^{\Lambda}p^2 dp
  =\frac{\Lambda^3L^3 }{6\pi^2},
\end{eqnarray}
where $\Lambda$ is the ultraviolent    cutoff in the energy of these modes. 
  This cutoff makes $N$  finite, and now the    Fock states can be
written in terms of the   occupying number $n_i$, 
\begin{eqnarray}
 \mid\Psi\!>=\mid n(\vec{k}_1), n(\vec{k}_2),\cdots,n(\vec{k}_N)>
 ~~\to~~ \mid n_1,n_2,\cdots,n_N>,
\end{eqnarray}
The dimension of the Hilbert space   can be  calculated from 
the  number of occupancies
 $\{n_i\}$. If no gravitational collapse occurs, we obtain  
\begin{eqnarray} \label{energy}
E= \sum_{i=1}^N n_i \omega_i  \leq
 E_{BH}=L.
\end{eqnarray}
Now the bound on the energy can be obtained as follows, 
\begin{eqnarray}
E \rightarrow
\frac{L^3}{2\pi^2}\int_{0}^{\Lambda} p^3 dp =\frac{\Lambda^4 L^3}{8\pi^2} \leq
 E_{BH}.
\end{eqnarray}
So, we can write $\Lambda^2\leq \frac{1}{L}.$ Now we can write the 
  maximum entropy as
\begin{eqnarray}
S_{\rm max}=-\sum_{j=1}^W \frac{1}{W}\ln\frac{1}{W}= \ln W\label{maxentropy},
\end{eqnarray}
where  $W$ is given by 
\begin{eqnarray}
W ={\rm dim}{\cal H}< \sum^N_{m=0}\frac{z^m}{(m!)^2} \leq
\sum^\infty_{m=0}\frac{z^m}{(m!)^2}=I_0(2\sqrt{z}) \sim
\frac{e^{2\sqrt{z}}}{\sqrt{4\pi\sqrt{z}}}.
\end{eqnarray}
Here $I_0$ is the zeroth-order Bessel function of the second kind.
As $z$ is given
by
\begin{eqnarray}
 z= \sum^N_{i=1}L_i \to  \frac{L^3}{2\pi^2}\int_{0}^{\Lambda}\Bigg[\frac{E_{BH}}{p}\Bigg]p^2dp
 =\frac{\Lambda^2L^4}{4\pi^2}, 
\end{eqnarray}
so we obtain, 
$
z  \leq  L^3.
$
We also have $A\sim L^2$, so we can write 
 the  Yurtsever's holographic bound on the maximum entropy  as  \cite{a,b,c,d}
\begin{eqnarray}
S_{\rm max}=\ln W \leq A^{3/4}.
\end{eqnarray}
This is  the Yurtsever's holographic bound on the  entropy which is obtained from  local quantum field theories. 

\section{Violation of the  Holographic Bound}
In this section, we will analyse the violation of the holographic principle from LQG. 
We again  consider a massless bosonic field confined to a cubic box of size $L$. However, now we will also 
consider the modifications to this analysis which stem from polymeric quantization. As the polymeric 
quantization is a characteristic feature of LQG, we will use this analysis to demonstrate the violation 
of the holographic principle in LQG.

Here, from this point on, we will drop sub index $\mu$ from all the momentum operators for notational clarity. 
Now using (\ref{Liou9}) and integrating over the position coordinates the total number of the quantized modes gets modified in the polymeric framework as 
\begin{eqnarray}\label{holo1}
  N\rightarrow \frac{L^{3}}{2\Pi^{2}}\int_{0}^{\lambda}\frac{p^{2}dp}{\sqrt{1-\mu^{2}p^{2}}}\;.
\end{eqnarray}
Now using the  condition $\frac{1}{\mu}\geq\lambda$, we obtain the following result
\begin{eqnarray}\label{holo2}
  N\approx\frac{L^{3}}{2\Pi^{2}}\Bigg(\int_{0}^{\lambda}p^{2}(1+\frac{1}{2}\mu^{2}p^{2})dp\Bigg)=\frac{L^{3}}{2\Pi^{2}}\Bigg(\frac{\lambda^{3}}{3}+\frac{\mu^{2}}{10}\lambda^{5}\Bigg).
\end{eqnarray}
The energy bound of the local field is modified to
\begin{eqnarray}\label{holo3}
  E\rightarrow \frac{L^{3}}{2\Pi^{2}}\int_{0}^{\lambda}\frac{p^{3}dp}{\sqrt{1-\mu^{2}p^{2}}}
  \approx \frac{L^{3}}{2\Pi^{2}}\Bigg(\frac{\lambda^{4}}{4}+\frac{\mu^{2}}{12}\lambda^{6}\Bigg)
 \leq  E_{BH}
\end{eqnarray}
where in the last step we have  used the fact that  black hole energy is the upper bound to the energy of the local field theory.
Using $E_{BH}=L$ and $\lambda^{2}\leq\frac{1}{L}$,   up to first order in $\mu^{2}$, we obtain the UV cut-off as
\begin{eqnarray}\label{holo5}
  \frac{L^{3}}{8\Pi^{2}}\Bigg(\lambda^{4}+\frac{\mu^{2}}{48}\lambda^{6}\Bigg)&\leq& L
 \nonumber \\ 
  \frac{1}{L}\Bigg(1-\frac{\mu^{2}}{96L}\Bigg) &\geq& \lambda^{2}
\end{eqnarray}
The modified entropy can now be  written as 
$S_{max}=lnW$   with $W\sim e^{2\sqrt{z}}$.
where 
\begin{eqnarray}\label{holo9}
  z\rightarrow \frac{L^{3}}{2\Pi^{2}}\int_{0}^{\lambda}\Big(\frac{E_{BH}}{p}\Big)
  \frac{p^{2}dp}{\sqrt{1-\mu^{2}p^{2}}}\
\approx
 \frac{L^{4}}{2\Pi^{2}}\Bigg(\frac{\lambda^{2}}{2}+\frac{\mu^{2}}{8}\lambda^{4}\Bigg)
\end{eqnarray}
From (\ref{holo5}) and (\ref{holo9}) the bound on $z$ is  
\begin{eqnarray}\label{holo11}
z\leq L^{3}\Bigg(1+\frac{5\mu^{2}}{96L}\Bigg)\:.
\end{eqnarray}
Thus, we can write the maximum entropy $ S_{max}= lnW=\ln e^{\sqrt{z}}=\sqrt{z}$ as follows, 
\begin{eqnarray}\label{holo17}
  S_{max} \leq L^{\frac{3}{2}}\Bigg(1+\frac{5\mu^{2}}{96L}\Bigg)^{\frac{1}{2}}
=A^{\frac{3}{4}}+\frac{5\mu^{2}}{192}\;A^{\frac{1}{4}}\:.
\end{eqnarray}  
Thus,  the polymer quantization violates
the Yurtsever's holographic bound on entropy  by a term which is of the order of $A^{\frac{1}{4}}$.
As the polymer quantization is the main characteristic technique used in loop quantum gravity, we expect that the loop quantum gravity 
will also violate the holographic principle. 

\section{Conclusion}
In this paper, we have analysed the effect of loop quantum gravity on the holographic principle. We have demonstrated that 
the Yurtsever's holographic  bound on entropy in the local quantum field theories 
is violated due to the effects coming from polymeric quantization. 
As the polymeric quantization is the basis of loop quantum gravity, our calculations indicate a violation of  the 
holographic principle in  
loop quantum gravity.  It would be interesting to analyse the violation of the 
holographic principle in a more general scene, as such a violation can have interesting physical 
consiquences. This is because various interesting physical systems have been analysed using the 
 general form of the holographic principle. 
The holographic principle has become the bases of the 
holographic cosmology \cite{cosm, c1, 1c, cosm1}.  The holographic cosmology is based on the 
idea that the difference between the degrees of 
freedom in a region  and the degrees of freedom on the boundary surrounding that region   
drives the expansion of the universe. This model is analysed using the  Jacobson formalism  \cite{4}. 
In this formalism, the   Einstein equations are  
viewed as the  Clausius relation. This is done by requiring 
the entropy   to be proportional to the area of the cosmological 
horizon. Thus,  the Friedmann
equations are obtained from  the Clausius relation using this formalism  \cite{zza1}.
It will be interesting to analyse the effect of the violation of the holographic principle on the 
holographic cosmology. It is expected that in this case, the speed of acceleration of the universe will 
get effected from the violation of the holographic principle. 
 The holographic principle has led to the development of the AdS/CFT correspondence, and 
 the AdS/CFT correspondence has in turn had many important applications \cite{ads, ads1}. Apart from 
 the conventional applications in the string theory \cite{st}, and M-theory \cite{m},
 the AdS/CFT correspondence has also 
 been used for understanding condensed matter systems \cite{cond, cond1}. 
 So, a violation of the holographic principle can  have interesting applications for all these systems. 
 
\section*{Acknowledgments}
We would like to thank Durmu{\c s} Ali Demir for useful discussions.


\begin{thebibliography}{99}

\bibitem{lqg1}A. Ashtekar, S. Fairhurst and J. Willis, Class. Quant. Grav. 20, 1031 (2003) 
\bibitem{lqg2}K. Fredenhagen and F. Reszewski, Class. Quantum Grav.  23, 6577 (2006) 
\bibitem{pq} A. Corichi, T. Vukasinac and J.A. Zapata, Class. Quant.  Grav. 24, 1495 (2007) 
\bibitem{pq1}A. Corichi, T. Vukasinac and J.A. Zapata, Phys. Rev. 76, 044016 (2007) 
\bibitem{pq3} D. A. Demir and O. Sarg{\i}n,   Phys. Lett. A 378,  3237 (2014) 
\bibitem{qf} G. M.  Hossain, V.  Husain and S. S. Seahra,   Phys. Rev. D82, 124032 (2010) 
\bibitem{in} S. M.  Hassan, V. Husain and S.  S. Seahra,  Phys. Rev. D 91, 065006 (2015) 
\bibitem{in12} G.  M.  Hossain, V. Husain and S.  S. Seahra, Phys. Rev. D81, 024005 (2010)  

\bibitem{holo}R.  Bousso, Rev. Mod. Phys. 74, 825 (2002)
\bibitem{holo1}G. t' Hooft,  gr-qc/9310026 


 \bibitem{z1}J. D. Bekenstein,  Phys. Rev. D 9, 3292 (1974)
 \bibitem{jd} J.  D. Bekenstein, Phys. Rev. D 7, 2333 (1973) 
 \bibitem{4a}N. Altamirano, D. Kubiznak, R. B. Mann, and Z. Sherkatghanad, Galaxies 2, 89 (2014)
\bibitem{z2}] S. W. Hawking, Nature 248, 30 (1974)





\bibitem{l1}S. Das, P. Majumdar and R. K. Bhaduri,  Class. Quant. Grav. 19, 2355 (2002)
\bibitem{card}T. R. Govindarajan, R. K. Kaul, V. Suneeta, Class. Quant. Grav. 18,  2877 (2001)
\bibitem{other} R. B. Mann and  S. N. Solodukhin, Nucl. Phys. {  B523},  293
(1998)
\bibitem{other0}
A. J. M. Medved and  G. Kunstatter, Phys. Rev. {  D60},   104029 (1999)
\bibitem{other1}
A. J. M. Medved and G. Kunstatter, Phys. Rev. {  D63}, 104005 (2001)
\bibitem{solo1} S. N. Solodukhin, Phys. Rev. {  D57},  2410 (1998)
\bibitem{solo2} A.  Sen,  JHEP  04, 156  (2013)
\bibitem{solo4} A.  Sen,  Entropy  13,  1305 (2011)
\bibitem{solo5} D. A. Lowe and S.  Roy,  Phys. Rev. D82, 063508 (2010)
\bibitem{jy} J. Jing and  M. L Yan, Phys. Rev. {  D63},  24003  (2001)
\bibitem{bss} D. Birmingham and S. Sen, Phys. Rev. {  D63}, 47501 (2001)


\bibitem{viol}S. Kalyana Rama, Phys. Lett. B 457, 268 (1999) 
\bibitem{viol1}D. Bak and S. J. Rey, Class. Quant. Grav. 17, L1 (2000)



\bibitem{a} U. Yurtsever, Phys. Rev. Lett. 91, 041302 (2003) 
\bibitem{b}A. Aste, Europhys. Lett. 69, 36 (2005) 
\bibitem{c} Y. X. Chen and Y. Xiao, Phys. Lett. B 662, 71 (2008)  
 \bibitem{d} Y. W. Kim, H. W.  Lee and  Y. S. Myung, Phys. Lett. B673, 293 (2009)


\bibitem{ali}A. F. Ali, Class. Quant. Grav. 28, 065013 (2011) 
\bibitem{ba} B. Majumder and S.  Sen,   Phys. Lett. B 717,  291 (2012) 


\bibitem{r1}R. Garattini, Phys.Lett. B 459, 461  (1999) 
\bibitem{r2}R. Garattini,  Int. J. Mod. Phys. A 17, 1965 (2002) 

\bibitem{Liou} M. A. Gorji, K. Nozari and B. Vakili, Class. Quant. Grav. 32, 155007 (2015)

\bibitem{cosm}T. Padmanabhan, Class. Quant. Grav. 21, 4485 (2004)
\bibitem{c1}T.~Padmanabhan and H.~Padmanabhan, Int.  J. Mod. Phys. D 23,  1430011 (2014)
\bibitem{1c}T.~Padmanabhan,  arXiv:1210.4174  
\bibitem{cosm1}T. Padmanabhan, arXiv:1206.4916 
\bibitem{4}J. Jacobson, Phys. Rev. Lett. 75, 1260 (1995)
\bibitem{zza1} R. G. Cai and S. P. Kim, JHEP 0502, 050 (2005)

 \bibitem{ads}J. M. Maldacena,   Adv. Theor. Math. Phys. 2, 231 (1998)
\bibitem{ads1}E. Witten,   Adv. Theor. Math. Phys. 2, 253 (1998)
\bibitem{st} O. Aharony, S. S. Gubser, J. Maldacena, H. Ooguri and Y. Oz, Phys. Rept. 323, 183 (2000)
\bibitem{m} O.  Aharony, O.  Bergman, D.  L.  Jafferis and  J.  Maldacena, JHEP 0810, 091 (2008)
\bibitem{cond} K.  Jensen,   JHEP 1101, 109 (2011) 
\bibitem{cond1}  A.  Donos, J.  P. Gauntlett, N.  Kim and O.  Varela,  JHEP,  1012, 003 (2010)  

  \end{thebibliography}
\end{document}